\documentclass[aps,pra,twocolumn,superscriptaddress]{revtex4}
\usepackage{amsfonts,amssymb,graphicx,amsmath}
\usepackage{amsmath}
\usepackage[usenames,dvipsnames]{xcolor}
\usepackage{pdfpages}
\usepackage{footmisc}
\usepackage[normalem]{ulem}
\usepackage{color}

\begin{document}

\title{Exact solutions of few-magnon problems in the spin-$S$ periodic XXZ chain}
\author{Ning Wu}
\email{wunwyz@gmail.com}
\affiliation{Center for Quantum Technology Research, School of Physics, Beijing Institute of Technology, Beijing 100081, China and Key Laboratory of Advanced Optoelectronic Quantum Architecture and Measurements (MOE), School of Physics, Beijing Institute of Technology, Beijing 100081, China}
\author{Hosho Katsura}
\email{katsura@phys.s.u-tokyo.ac.jp}
\affiliation{Department of Physics, The University of Tokyo, 7-3-1 Hongo, Bunkyo-ku, Tokyo 113-0033, Japan}
\affiliation{Institute for Physics of Intelligence, The University of Tokyo, 7-3-1 Hongo, Bunkyo-ku, Tokyo 113-0033, Japan}
\affiliation{Trans-Scale Quantum Science Institute, University of Tokyo, Bunkyo-ku, Tokyo 113-0033, Japan}
\author{Sheng-Wen Li}
\affiliation{Center for Quantum Technology Research, School of Physics, Beijing Institute of Technology, Beijing 100081, China and Key Laboratory of Advanced Optoelectronic Quantum Architecture and Measurements (MOE), School of Physics, Beijing Institute of Technology, Beijing 100081, China}
\author{Xiaoming Cai}
\affiliation{State Key Laboratory of Magnetic Resonance and Atomic and Molecular Physics, Wuhan Institute of Physics and Mathematics, APM, Chinese Academy of Sciences, Wuhan 430071, China}
\author{Xi-Wen Guan}
\email{xwe105@wipm.ac.cn}
\affiliation{State Key Laboratory of Magnetic Resonance and Atomic and Molecular Physics, Wuhan Institute of Physics and Mathematics, APM, Chinese Academy of Sciences, Wuhan 430071, China}
\affiliation{Center for Cold Atom Physics, Chinese Academy of Sciences, Wuhan 430071, China}
\affiliation{Department of Theoretical Physics, Research School of Physics and Engineering, Australian National University, Canberra ACT 0200, Australia}
\begin{abstract}
We solve few-magnon problems for a \emph{finite-size} spin-$S$ periodic Heisenberg XXZ chain with single-ion anisotropy through constructing sets of \emph{exact} Bloch states achieving block diagonalization of the system. Concretely, the two-magnon (three-magnon) problem is converted to a single-particle one on a one-dimensional (two-dimensional) effective lattice whose size depends linearly (quadratically) on the total number of sites. For parameters lying within certain ranges, various types of multimagnon bound states are manifested and shown to correspond to edge states on the effective lattices. In the absence of the single-ion anisotropy, we reveal the condition under which exact zero-energy states emerge. As applications of the formalism, we calculate the transverse dynamic structure factor for a higher-spin chain near saturation magnetization and find signatures of the multimagnon bound states. We also calculate the real-time three-magnon dynamics from certain localized states, which are relevant to cold-atom quantum simulations, by simulating single-particle quantum walks on the effective lattices. This provides a physically transparent interpretation of the observed dynamics in terms of propagation of bound state excitations. Our method can be directly applied to more general spin or itinerant particle systems possessing translational symmetry.
\end{abstract}

\maketitle
\section{Introduction}
\par The Heisenberg XXZ model is a paradigmatic model exhibiting strong correlations. On one hand, dynamical properties of the spin-1/2 XXZ chain continue to attract the attention of the solid-state- and mathematical-physics communities~\cite{Balents,Bethestring,Chauhan2020,Suzuki}. On the other hand, recent experimental advances in cold-atom systems enable realizations of the XXZ chain and preparation of certain initial states~\cite{Fukuhara2013,Nature2020}, even with higher spins~\cite{Ketterle2021}, providing an ideal setting for studying nonequilibrium quantum dynamics. Recently, few-magnon dynamics in the spin-1/2 and spin-1 XXZ chains has also attracted great theoretical interest~\cite{Essler2012,Andrei,HJC2021}. Magnons (or spin waves) are elementary excitations in the saturated regime of quantum magnetic systems and play an important role in understanding magnetism, magnetic order, and spin dynamics, etc. In particular, multimagnon bound states, which were first predicted by Bethe in studying the spin-1/2 Heisenberg chain, are believed to be difficult to detect experimentally, though evidence of few-magnon bound states has been revealed in spin-ladder systems through spectroscopic studies~\cite{Spinladder1,Spinladder2}. It was theoretically proposed~\cite{Essler2012}, and later experimentally verified~\cite{Fukuhara2013}, that these magnon bound states can be observed using multimagnon quantum walks. It was shown recently that the appearance of multimagnon bound states in an antiferromagnetic spin-1/2 chain can also be uncovered in the transverse dynamic structure factor~\cite{Balents}.
\par As a theoretical problem, the few-magnon physics in higher-spin Heisenberg-like models has long been studied by a variety of approaches, including Green's function~\cite{Wortis1963,Hanus1973,Loly1986}, the Dyson-Maleev transformation~\cite{Silberglitt1970,JPSJ1971,Tjon1974,PRB1981}, Bethe ansatz~\cite{Papan1987,Bibikov2016}, continuous unitary transformations~\cite{PRB2004}, and center-of-mass analysis~\cite{Hanus1963,Torrance1969,Shouthern1989,Shouthern1989PRB,Southern1994,Southern1996,Southern1998,Furusaki2007,Lee2017}, etc. Among these, the center-of-mass method provides a physically intuitive way to convert the few-magnon problem into a single-particle one~\cite{Torrance1969,Southern1994,Furusaki2007}. In an early work, Southern, Lee, and Lavis studied the nature of three-magnon excitations in general infinite-size spin-$S$ chains by constructing a set of Bloch states forming a semi-infinite triangle-shape effective lattice~\cite{Southern1994}. Kecke, Momoi, and Furusaki used similar ideas to study the emergence of multimagnon bound states in infinite-size frustrated ferromagnetic chains~\cite{Furusaki2007}. Nevertheless, in experimentally relevant cases the spin system of interest always has a finite number of sites. It is therefore important and demanding to find out an exact set of Bloch basis states (for finite chains) that can form a finite-size effective lattice. To the best of our knowledge, such a mathematically rigorous treatment of the three-magnon problem for a \emph{finite-size} higher-spin XXZ chain is still missing.
\par In this paper, we construct exact Bloch states achieving a block diagonalization of the two- and three-magnon sectors in a finite-size spin-$S$ XXZ chain with single-ion anisotropy. This converts the two-magnon (three-magnon) problem into a single-particle one on a one-dimensional (two-dimensional) effective lattice whose size scales linearly (quadratically) with the total number of sites. Our method provides an exact, intuitive, and convenient way to understand the few-magnon physics.
\par We employ our formalism to study several aspects of the model. We first reveal the condition under which the few-magnon excitation energy with respect to the ferromagnetic state exactly vanishes and obtain explicit forms of these zero-energy states as certain Bloch states. These states are intimately related to the spin helix states, which have recently attracted much theoretical~\cite{Helix,PRB2021} and experimental~\cite{Nature2020,Ketterle2021} attention. In certain parameter regimes, we reveal various types of multimagnon bound states, which turn out to be localized edge states on the effective lattice.
\par We then turn to study the dynamical properties of the system. We extend the analysis of the transverse dynamic structure factors in Ref.~\cite{Balents} for spin-1/2 chains in the high magnetization regime to the case of higher spins. In an early work, Silberglitt and Torrance showed that~\cite{Silberglitt1970} for $S>1/2$ the so-called single-ion two-magnon bound states, which correspond to two spin derivations on the same site, might emerge besides the usual exchange two-magnon bound states (corresponding to two spin derivations on two nearest-neighboring sites). We show that for higher spins the appearance of both the usual exchange and the single-ion (unique for $S>1/2$) two-magnon bound states can be uncovered in the experimentally accessible transverse dynamic structure factor. Similarly, the appearance of three-magnon bound states can also be uncovered in the transverse dynamic structure factor for two-magnon eigenstates, which involves the transition between the two- and three-magnon sectors. As another dynamical application of our formalism, we calculate the three-magnon dynamics from localized spin states via simulating single-particle quantum walks on the effective lattices. We use several perturbative approaches, including the degenerate many-body perturbation and the time-dependent perturbation theory, to interpret the obtained three-magnon spectra and three-magnon dynamics and to demonstrate the essential role played by the three-magnon bound states in the magnetization diffusions.
\par The rest of the paper is organized as follows. In Sec.~\ref{SecII}, we introduce the one-dimensional spin-$S$ XXZ model with single-ion anisotropy and present in detail the construction of exact Bloch states in the two- and three-magnon sectors. In Sec.~\ref{SecIII}, we study the emergence of zero-energy states and find out the relationship between these states and certain eigenstates in the Bloch space. In Sec.~\ref{SecIV}, we present detailed numerical results for the two-magnon sector, including the two-magnon excitation spectrum, the two-magnon bound states and their wave functions in the Bloch space, and the dynamic structure factor near saturation magnetization. In Sec.~\ref{SecV}, we study in detail the three-magnon bound states and three-magnon dynamics. Conclusions are drawn in Sec.~\ref{SecVI}.
\section{Model and methodology}\label{SecII}
\subsection{Model}
The spin-$S$ XXZ chain with $N$ spins is described by the Hamiltonian
\begin{eqnarray}\label{Haml}
H&=&- J_{xy}H_{XY}-J_zH_Z-DH_D,\nonumber\\
H_{XY}&=&\sum^N_{j=1}(S^x_jS^x_{j+1}+S^y_jS^y_{j+1}),\nonumber\\
H_{Z}&=&\sum^N_{j=1}S^z_jS^z_{j+1},~~H_{D}=\sum^N_{j=1}(S^z_j)^2,
\end{eqnarray}
where $\vec{S}_j=(S^x_j,S^y_j,S^z_j)$ is the spin operator on site $j$ with quantum number $S\geq1/2$, $J_{xy}$ and $J_z$ are the exchange interactions between nearest-neighboring spins, and $D$ is the single-ion anisotropy strength. It is easy to see that the total magnetization $M=\sum_jS^z_j$ is conserved.
\par We assume that $N$ is even and impose the periodic boundary condition $\vec{S}_j=\vec{S}_{N+j}$, which guarantees the translational invariance of the chain. Unless otherwise specified, we focus on the case of $J_z>0$ and take the ferromagnetic state $|F\rangle=|S,S,\cdots,S\rangle$ as a reference state possessing eigenenergy $E_F=-NS^2(J_z+D)$, though our formalism is valid for both a ferromagnetic chain and an antiferromagnetic chain (with $J_z<0$) near saturation magnetization~\cite{Balents} (see Sec.~\ref{SecIVB} below).
\par  The $n$-magnon sector is defined as the subspace spanned by all the spin configurations having magnetization $NS-n$,
\begin{eqnarray}\label{jjj}
|j_1,j_2,\cdots, j_n\rangle\equiv CS^-_{j_1}S^-_{j_2}\cdots S^-_{j_n}|F\rangle,
\end{eqnarray}
where $C$ is a suitable normalization coefficient and the site indices $1\leq j_1\leq j_2\leq\cdots \leq j_n\leq N$ are not necessarily distinct for $S>1/2$. We define the translation operator $T$ by
\begin{eqnarray}
T|j_1,j_2,\cdots, j_n\rangle=|j_1+1,j_2+1,\cdots, j_n+1\rangle.
\end{eqnarray}
\par The $N$ one-magnon states are simply
\begin{eqnarray}\label{psik1}
|\psi(k)\rangle=\frac{1}{\sqrt{N}}\sum^{N-1}_{n=0}e^{ikn}T^n|1\rangle,~k\in K_0,
\end{eqnarray}
where the wave number $k$ lives in the set
\begin{eqnarray}\label{K0}
K_0=\left\{-\pi,-\pi+\frac{2\pi}{N},\cdots,0,\cdots,\pi-\frac{2\pi}{N}\right\},
\end{eqnarray}
which ensures the translational invariance of $|\psi(k)\rangle$, i.e., $T|\psi(k)\rangle=e^{-ik}|\psi(k)\rangle$. The one-magnon state $|\psi(k)\rangle$ is itself an eigenstate of $H$ with eigenenergy $E_F+\mathcal{E}_1(k)$, where $\mathcal{E}_1(k)=2S(J_z-J_{xy}\cos k)+D(2S-1)$.
\subsection{Two-magnon sector}
\par In this subsection, we assume $S\geq 1$ since the case of $S=1/2$ can be obtained as a limiting case of the formalism developed below. In the two-magnon sector, two types of real-space basis states,
\begin{eqnarray}
|i,j\rangle&=&\frac{1}{2S}S^-_iS^-_j|F\rangle,~~i<j,
\end{eqnarray}
and
\begin{eqnarray}
|i,i\rangle&=&\frac{1}{2\sqrt{S(2S-1)}} (S^-_i)^2|F\rangle,
\end{eqnarray}
are allowed for $S>1/2$~\cite{Papan1987}. These $N(N+1)/2$ basis states can be obtained by successively applying the translation operator $T$ to the $N/2+1$ \emph{parent states}, $|1,1\rangle,|1,2\rangle,\cdots,|1,N/2\rangle,~\mathrm{and}~|1,N/2+1\rangle$. Among these, $|1,j\rangle~(j=1,2,\cdots,N/2)$ generates $N-1$ additional states under the action of $T$, while the special state $|1,N/2+1\rangle$ generates only $N/2-1$ additional states.
\par These observations suggest that we need to construct two different types of Bloch states,
\begin{eqnarray}\label{Bloch2m1}
|\psi_r(k)\rangle&=&\frac{e^{i\frac{rk}{2}}}{\sqrt{N}}\sum^{N-1}_{n=0}e^{ikn}T^n|1,1+r\rangle,~r=0,\cdots,\frac{N}{2}-1,\nonumber\\
\end{eqnarray}
and
\begin{eqnarray}\label{Bloch2m2}
|\psi_{\frac{N}{2}}(k)\rangle&=&e^{i\frac{Nk}{4}}\sqrt{\frac{2}{N}}\sum^{N/2-1}_{n=0}e^{ikn}T^n|1,1+\frac{N}{2}\rangle,
\end{eqnarray}
where $r$ measures the relative distance between the two down spins in a parent state and the factors $e^{i\frac{rk}{2}}$ and $e^{i\frac{Nk}{4}}$ are introduced for later convenience~\cite{Furusaki2007}.
\par It is easy to check that for any $k\in K_0$ we have $T|\psi_r(k)\rangle=e^{-ik}|\psi_r(k)\rangle$ ($r=0,1,\cdots,N/2-1$). However, the property $T|\psi_{\frac{N}{2}}(k)\rangle=e^{-ik}|\psi_{\frac{N}{2}} (k)\rangle$ holds only if $e^{ikN/2}=1$, which restricts the allowed wave numbers to a subset $K_1$ of $K_0$, i.e.,
\begin{eqnarray}\label{K1e}
K_1=\left\{-\pi,-\pi+\frac{4\pi}{N},\cdots,0,\cdots,\pi-\frac{4\pi}{N}\right\},
\end{eqnarray}
for even $\frac{N}{2}$, or
\begin{eqnarray}\label{K1o}
K_1=\left\{-\pi+\frac{2\pi}{N},-\pi+\frac{6\pi}{N},\cdots,0,\cdots,\pi-\frac{2\pi}{N}\right\},
\end{eqnarray}
for odd $\frac{N}{2}$.
\par Conversely, the local state $|\phi^n_r\rangle\equiv T^n|1,1+r\rangle$ can be expanded in terms of the Bloch states as
\begin{eqnarray}\label{phinr}
|\phi^n_r\rangle=\frac{1}{\sqrt{N}}\sum_{k\in K_0}e^{-ikn-i\frac{kr}{2}}|\psi_r(k)\rangle
\end{eqnarray}
for $r=0,1,\cdots,\frac{N}{2}-1$, and
\begin{eqnarray}\label{phinN}
|\phi^n_{\frac{N}{2}}\rangle=\sqrt{\frac{2}{N}}\sum_{k\in K_1}e^{-ikn-i\frac{kN}{4}}|\psi_{\frac{N}{2}}(k)\rangle
\end{eqnarray}
for $r=N/2$.
\par We denote the complement of $K_1$ as $K'_1$, so that $K_0=K_1\bigcup K'_1$. For each $k\in K_1$, we find after some straightforward calculation that the $\frac{N}{2}+1$ ordered Bloch states $\{|\psi_0(k)\rangle,|\psi_1(k)\rangle,\cdots,|\psi_{\frac{N}{2}}(k)\rangle\}$ form a closed basis and result in the tridiagonal block Bloch Hamiltonian,
\begin{widetext}
\begin{eqnarray}\label{Hk1}
\mathcal{H}_2(k)=E_F +\left(
      \begin{array}{cccccccc}
        \Omega_0 &   \sqrt{S(2S-1)}A_k &   & &    &    &   &   \\
          \sqrt{S(2S-1)}A_k & \Omega_1 &  SA_k &  &   &   &   &   \\
          &  SA_k & \Omega_2 &  SA_k  & &   &   &   \\
          &   &  SA_k &  \Omega_2 &  &    &   &   \\
           &   &    &  & \ddots&   &   &   \\
          &   &   &     & &\Omega_2 &  SA_k &   \\
          &   &   &    &  &SA_k & \Omega_2 &   \sqrt{2} SA_k \\
          &   &   &    & &  &   \sqrt{2} SA_k & \Omega_2 \\
      \end{array}
    \right),~~k\in K_1,
\end{eqnarray}
\end{widetext}
where
\begin{eqnarray}
A_k& \equiv & -2J_{xy}\cos\frac{k}{2},\nonumber\\
\Omega_0 & \equiv &  4 S J_z +4(S-1)D,\nonumber\\
\Omega_1 & \equiv &  (4S-1)J_z+2(2S-1)D,\nonumber\\
\Omega_2 & \equiv &  4 S J_z +2(2S-1)D.
\end{eqnarray}
It is worth noting that $\Omega_2\geq \Omega_0, \Omega_1$ for $J_z,D\geq 0$. For $k\in K'_1$, the Bloch state $|\psi_{\frac{N}{2}}(k)\rangle$ is not properly defined. The Bloch Hamiltonian $\mathcal{H}_2(k\in K'_1)$ can therefore be obtained by eliminating the last row and the last column from $\mathcal{H}_2(k\in K_1)$.
\par
\begin{figure}
\includegraphics[width=.50\textwidth]{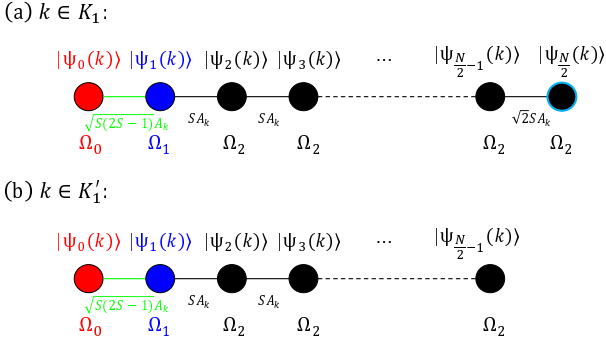}
\caption{The effective one-dimensional lattice formed by the Bloch states $\{|\psi_r(k)\rangle\}$ in the two-magnon sector with wave numbers drawn from (a) $k\in K_1$ and (b) $k\in K'_1$.}
\label{Fig1D}
\end{figure}

 Physically, we can view $\mathcal{H}_2(k\in K_1)$ [$\mathcal{H}_2(k\in K'_1)$] as a single-particle problem on an effective one-dimensional lattice with $ N/2+1$ ($N/2$) sites, with the nearest-neighboring hopping proportional to $A_k$ and the on-site energies being $\Omega_i$ (see Fig.~\ref{Fig1D}).
The completeness of the Bloch basis can be verified by noting that $\frac{N}{2}(\frac{N}{2}+1)+\frac{N}{2}\frac{N}{2}=\frac{1}{2}N(N+1)$.
\subsection{Three-magnon sector}
\par We now turn to discuss the more subtle three-magnon sector. In this subsection, we assume $S\geq3/2$ and $N=3m$ (hence $m=$ even). As we will see, the cases of $N=3l\pm 1~(l\in \mathbb{Z})$ can be analyzed in a similar but simpler way. It is obvious that all the real-space basis states can be classified into the following three types:
\begin{eqnarray}\label{jjj}
&&\mathrm{(i)}~|j_1,j_1,j_1\rangle,~1\leq j_1\leq N,\nonumber\\
&&\mathrm{(ii)}~|j_1,j_1,j_2\rangle~\mathrm{and}~|j_1,j_2,j_2\rangle,~1\leq j_1<j_2\leq N,\nonumber\\
&&\mathrm{(iii)}~|j_1,j_2,j_3\rangle,~1\leq j_1<j_2<j_3\leq N.
\end{eqnarray}
These states form a complete basis of the $\mathcal{D}$-dimensional three-magnon sector, where $
\mathcal{D}=\binom{N}{1}+2\binom{N}{2}+\binom{N}{3}
=\frac{1}{6}N(N+1)(N+2)$.
\begin{figure*}
\includegraphics[width=.8\textwidth]{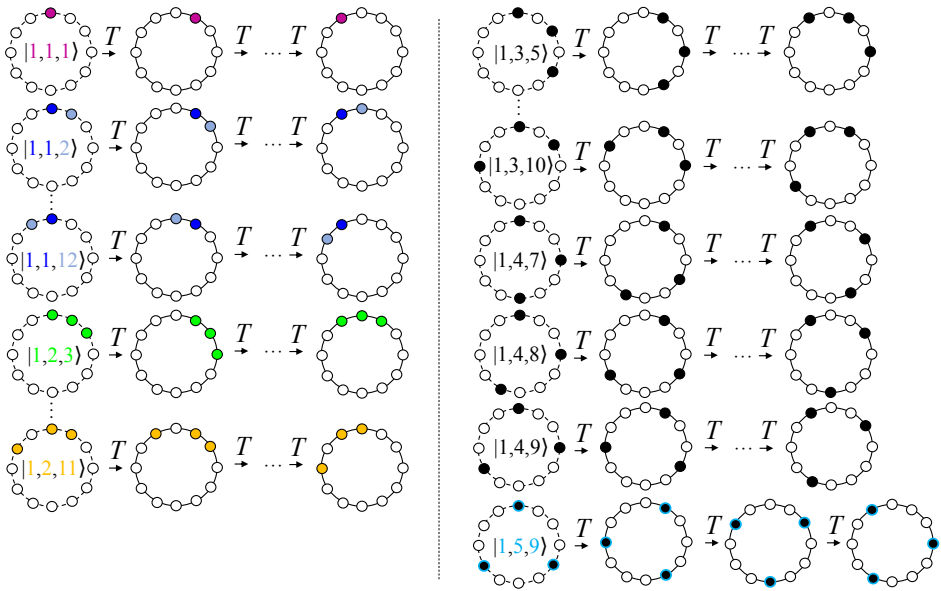}
\caption{The 364 real-space basis states in the three-magnon sector for $N=12$ and $S\geq\frac{3}{2}$. The leftmost dashed circles indicate the 31 parent states: $|1,1+r_1,1+r_1+r_2\rangle$ with $r_1=0,1,2,3$ and $r_2=r_1,r_1+1,\cdots,11-2r_1$, as well as a special one, $|1,5,9\rangle$. Note that $|1,5,9\rangle$ generates only three new states and no such special state exists for $N=3l\pm 1,~l\in \mathbb{Z}$.}
\label{3magnonreal}
\end{figure*}
\par To construct the Bloch states from the typical parent states $|\phi^0_{r_1,r_2}\rangle\equiv |1,1+r_1,1+r_1+r_2\rangle$ and their translations $\{|\phi^n_{r_1,r_2}\rangle\equiv T^n|\phi^0_{r_1,r_2}\rangle\}$, we need to further classify the $\mathcal{D}$ states given by Eq.~(\ref{jjj}) into groups having fixed $r_1$ and $r_2$. For example, the $N$ type $\mathrm{(i)}$ states are simply $|\phi^0_{0,0}\rangle,~|\phi^1_{0,0}\rangle,\cdots,|\phi^{N-1}_{0,0}\rangle$ (Fig.~\ref{3magnonreal}; left column, first row). The $N(N-1)$ type $(\mathrm{ii})$ states can be written as $|\phi^0_{0,r_2}\rangle,~|\phi^1_{0,r_2}\rangle,\cdots,|\phi^{N-1}_{0,r_2}\rangle$ with $1\leq r_2\leq N-1$ (Fig.~\ref{3magnonreal}; left column, row 2 to row $N$).
\par However, the $\binom{N}{3}$ type $\mathrm{(iii)}$ states need to be treated more carefully. As realized in an early work by Torrance and Tinkham~\cite{Torrance1969}, there exist ``complicated restrictions" on the $r_1$ and $r_2$ appearing in the parent state $|1,1+r_1,1+r_1+r_2\rangle$. The three excited sites in the state $|j_1,j_2,j_3\rangle$ divide the ring into three successive segments (ordered clockwise, see Fig.~\ref{3magnonreal} for examples) having lengths $j_2-j_1$, $j_3-j_2$, and $N-(j_3-j_1)$. To avoid double counting, we choose $r_1$ in $|1,1+r_1,1+r_1+r_2\rangle$ as
\begin{eqnarray}
r_1=\min\{j_2-j_1,j_3-j_2,N-(j_3-j_1)\},\nonumber
\end{eqnarray}
so that $r_1\leq r_2$ and $r_1\leq N-(r_1+r_2)$, giving $r_1\leq\frac{N}{3}=m$ and $r_1\leq r_2\leq N-2r_1$ for fixed $r_1$. Unless $r_1=m$, the two states with $r_2=r_1$ and $r_2=N-2r_1$ are connected by translations, and by choosing $r_2=r_1$ we have $r_1\leq r_2\leq N-(2r_1+1)$ for any $0\leq r_1<m$. For $r_1=m$ we must have $r_2=m$, giving the unique parent state $|\phi^0_{m,m}\rangle$, which is a three-magnon counterpart of the two-magnon parent state $|\phi^0_{\frac{N}{2}}\rangle$. Note that no such special states exist for $N=3l\pm 1$, $l\in \mathbb{Z}$. We now obtain all the $\bar{\mathcal{D}}+1$ parent states, where $\bar{\mathcal{D}}=\sum^{m-1}_{r_1=0}(N-3r_1)=\frac{1}{6}N(N+3)$.
\par Since for $r_1< m$ ($r_1=m$) the allowed parent state $|\phi^0_{r_1,r_2}\rangle$ generates $N-1$ ($m-1$) additional translated states, the total number of such obtained basis states is $N\cdot\bar{\mathcal{D}}+m\cdot 1=\mathcal{D}$, yielding a consistency. We thus complete the classification of the desired parent states and their translations that will be used to construct the Bloch states.
\par For $k\in K_0$ and for each pair of $(r_1,r_2)$ with $r_1<m$, we define the translationally invariant state~\cite{Furusaki2007}
\begin{eqnarray}\label{psir1r2}
|\psi_{r_1,r_2}(k)\rangle&=&\frac{e^{r_1i\frac{k}{3}}e^{(r_1+r_2)i\frac{k}{3}}}{\sqrt{N}}\sum^{N-1}_{n=0}e^{ikn}|\phi^{n}_{r_1,r_2}\rangle.
\end{eqnarray}
However, for the $m$ states $\{|\phi^n_{m,m}\rangle|0\leq n\leq m-1\}$ with $C_3$ symmetry, we have to construct the Bloch state as
\begin{eqnarray}\label{psimm}
|\psi_{m,m}(k)\rangle&=&\frac{e^{ikm} }{\sqrt{m}} \sum^{m-1}_{n=0}e^{ikn}|\phi^{n}_{m,m}\rangle.
\end{eqnarray}
To ensure the translational invariance of $|\psi_{m,m}(k)\rangle$, the wave number $k$ in the above equation must take value from the subset
\begin{eqnarray}\label{K2}
K_2=\left\{\frac{2\pi l}{m}\bigg| l=-\frac{m}{2},-\frac{m}{2}+1,\cdots,\frac{m}{2}-1\right\}.
\end{eqnarray}
Since there are $N$ ($m$) elements in $K_0$ ($K_2$), the total number of the Bloch states given by Eqs.~(\ref{psir1r2}) and (\ref{psimm}) is still $N\bar{\mathcal{D}}+m=\mathcal{D}$. We define the complement of $K_2$ as $K'_2=K_0\setminus K_2$, so that $e^{ikm}=1$ ($e^{ikm}\neq1$) for $k\in K_2$ ($k\in K'_2$).
\par A local state $|\phi^n_{r_1,r_2}\rangle$ can be expanded in terms of the Bloch states as
\begin{eqnarray}\label{phinrrexp}
|\phi^n_{r_1,r_2}\rangle&=&\sum_{k\in K_0}\frac{e^{-ikn}e^{-i\frac{k}{3}(2r_1+r_2)}}{\sqrt{N}}|\psi_{r_1,r_2}(k)\rangle
\end{eqnarray}
for $(r_1,r_2)\neq(m,m)$, and
\begin{eqnarray}\label{phinmmexp}
|\phi^n_{m,m}\rangle&=&\sum_{k\in K_2}\frac{e^{-ikn}e^{-ikm}}{\sqrt{m}}|\psi_{m,m}(k)\rangle.
\end{eqnarray}
In Sec.~\ref{SecV}, we will use Eqs.~(\ref{phinrrexp}) and (\ref{phinmmexp}) to calculate the three-magnon quantum walks in the Bloch space.
\begin{figure*}
\includegraphics[width=.9\textwidth]{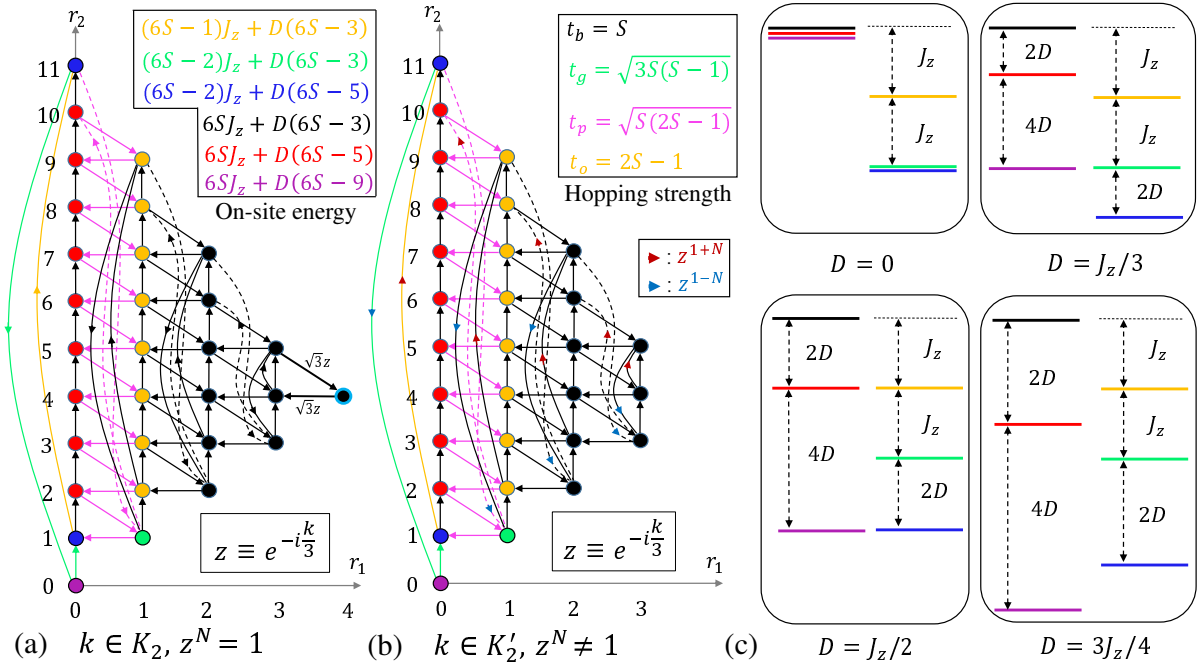}
\caption{Representation of $H_{XY}$ on an effective lattice in the $r_1$-$r_2$ plane formed by the Bloch basis states $\{|\psi_{r_1,r_2}(k)\rangle\}$ for (a) $k\in K_2$ and (b) $k\in K'_2$ ($N=12$ as an example). The colors of the circles indicate different eigenenergies of $\mathcal{H}_3(k)|_{J_{xy}=0}-E_F$ [shown in panel (a)]. Nonvanishing complex hopping between two Bloch states is represented by an arrowed line, with the color and arrow indicating its magnitude [shown in panel (b)] and phase factor, respectively. The action of $H_{XY}$ on a certain Bloch state can directly be read off. For example, for $k\in K'_2$ we have $H_{XY}|\psi_{1,1}(k)\rangle=t_pz|\psi_{0,1}(k)\rangle+t_pz^*|\psi_{0,2}(k)\rangle+t_pz^{N-1}|\psi_{0,11}(k)\rangle+t_pz^{1+N}|\psi_{0,10}(k)\rangle+t_bz^{N-1}|\psi_{1,9}(k)\rangle+t_bz|\psi_{1,2}(k)\rangle$. (c) Evolution of the on-site energies with respect to varying $J_z$ and $D$.}
\label{3magnon2D}
\end{figure*}
\par After a lengthy but straightforward calculation, we find that for each $k\in K_2$ the $\bar{\mathcal{D}}+1$ Bloch states $\{|\psi_{r_1,r_2\neq (m,m)}(k)\rangle\}$ and $|\psi_{m,m}(k)\rangle$ form a closed set under the action of the Hamiltonian $H$. This results in a $(\bar{\mathcal{D}}+1)$-dimensional Bloch Hamiltonian $\mathcal{H}_3(k\in K_2)$, which describes a single-particle problem on a triangle-shape effective lattice in the $r_1$-$r_2$ plane. It is apparent that the term $-J_zH_Z-DH_D$ is diagonal in the Bloch basis and serves as the on-site energy for the effective lattice, while the spin-flipping term $-J_{xy}H_{XY}$ contributes to the hopping among the lattice sites; see Fig.~\ref{3magnon2D}(a) for a detailed structure of the effective lattice (with $N=12$). For $k\in K'_2$, it can be similarly shown that the $\bar{\mathcal{D}}$ Bloch states $\{|\psi_{r_1,r_2\neq (m,m)}(k)\rangle\}$ form a closed set and yield a $\bar{\mathcal{D}}$-dimensional Bloch Hamiltonian $\mathcal{H}_3(k\in K'_2)$. The effective lattice corresponding to $\mathcal{H}_3(k\in K'_2)$ is shown in Fig.~\ref{3magnon2D}(b), where the site $(m,m)$ has been removed.
\par We now turn to some remarks regarding the three-magnon effective lattice. (i) The effective lattices for $S=1/2$ ($S=1$) can simply be obtained by removing the leftmost column [the $(0,0)$ site] of the original lattices. (ii) Compared to the two-magnon effective lattice where the nearest-neighbor hopping is real, in the three-magnon effective lattice there exist complicated long-range hopping terms and the hopping amplitudes are generally complex. (iii) Similar ideas have been developed in Refs.~\cite{Southern1994} and  \cite{Furusaki2007} for infinite chains. However, to the best of our knowledge, the exact Bloch states given by Eqs.~(\ref{psir1r2}) and (\ref{psimm}) provide the first mathematically rigorous construction of the Bloch basis in the three-magnon sector for a finite-size higher-spin XXZ chain. (iv) The obtained exact block Bloch Hamiltonians provide a convenient way to calculate dynamical properties of the system in momentum space.
\section{Exact zero-energy states for $D=0$}\label{SecIII}
\par As the first application of our formalism, let us study the emergence of exact zero-energy (with respect to the ferromagnetic state) multimagnon states. We will demonstrate the relationship between these zero-energy states and certain eigenstates in the Bloch space. In this section we focus on the case of $D=0$.
\par  For $D=0$, the one-magnon excitation energy $\mathcal{E}_1(k)$ vanishes when the following condition is satisfied,
\begin{eqnarray}\label{JzJxy}
J_z=J_{xy}\cos k,~k\in K_0.
\end{eqnarray}
The corresponding (unnormalized) one-magnon state is given by $L_k|F\rangle$, where $L_k\equiv\sum^N_{j=1}e^{ikj}S^-_j$ is a collective lowering operator.
\par It can be shown that (see Appendix~\ref{AppA}) $(L_k)^n|F\rangle$ is indeed a zero-energy state in the $n$-magnon sector once Eq.~(\ref{JzJxy}) is fulfilled, i.e.,
\begin{eqnarray}\label{ZEST}
(H-E_F)(L_k)^n|F\rangle=0,~n\leq 2NS+1.
\end{eqnarray}
Note that for $n>2NS+1$ we always have $(L_k)^n|F\rangle=0$. A direct corollary of Eq.~(\ref{ZEST}) is
\begin{eqnarray}
(L_k)^{2NS+1}|F\rangle\propto |-S,-S,\cdots,-S\rangle.
\end{eqnarray}
These zero-energy states are interesting since linear combinations of them are relevant to the so-called spin helix state~\cite{Helix,PRB2021}, which has recently been experimentally prepared for $S=1/2$~\cite{Nature2020,Ketterle2021}.
\par As an eigenstate of $H$, $(L_k)^n|F\rangle$ under the condition given by Eq.~(\ref{JzJxy}) must also be a zero-energy eigenstate of a certain Bloch Hamiltonian $\mathcal{H}_n(p)-E_F$, with $p=p(k)$ a function of $k$ to be determined. From the relation $TS^-_jT^{-1}=S^-_{j+1}$ we have $
T(L_k)^n|F\rangle=e^{-ink}(L_k)^n|F\rangle$, which means that $(L_k)^n|F\rangle$ possesses momentum $nk$. We therefore expect that $(L_k)^n|F\rangle$ is also the zero-energy eigenstate of $\mathcal{H}_n(p(k))-E_F$, where $p(k)$ is given by
\begin{eqnarray}\label{pkn}
p(k)=nk~(\mathrm{mod}~2\pi),
\end{eqnarray}
with the understanding that $p(k)\in[-\pi,\pi)$.
\par For $n=1$ it is easy to see that $p(k)=k$. For $n=2$, we can explicitly show that $p(k)=2k~(\mathrm{mod}~2\pi)$. Let us focus on the left half of the Brillouin zone, i.e., $k\in [-\pi,0]$ since the condition given by Eq.~(\ref{JzJxy}) is symmetric under $k\to -k$. For simplicity, we also consider the case of $N=4l$, $l\in \mathbb{Z}$. A straightforward calculation gives
\begin{widetext}
\begin{eqnarray}\label{Lk2b}
L^2_{k}|F\rangle&=&4S \sqrt{N} e^{i2k}\left[\frac{ \sqrt{ 2S-1 } }{2\sqrt{S}  }|\psi_0(2k+2\pi)\rangle+ \sum^{N/2-1}_{r=1}(-1)^r|\psi_r(2k+2\pi)\rangle+(-1)^{N/2}\frac{1}{\sqrt{2}}|\psi_{\frac{N}{2}}(2k+2\pi)\rangle\right]
\end{eqnarray}
for $k\in\{-\pi,-\pi+\frac{2\pi}{N},\cdots,-\frac{\pi}{2}-\frac{2\pi}{N}\}$, and
\begin{eqnarray}\label{Lk2a}
 L^2_{k}|F\rangle&=&  4S \sqrt{N} e^{i2k}\left[\frac{ \sqrt{ 2S-1 } }{2\sqrt{S}  }|\psi_0(2k)\rangle+ \sum^{N/2-1}_{r=1}|\psi_r(2k)\rangle+\frac{1}{\sqrt{2}}|\psi_{\frac{N}{2}}(2k)\rangle\right]
\end{eqnarray}
for $k\in\{-\frac{\pi}{2},-\frac{\pi}{2}+\frac{2\pi}{N},\cdots,0\}$.
\par On the other hand, the relation $p(k)=2k~(\mathrm{mod}~2\pi)$ gives $p\in K_1$ and $\cos k=\pm\cos\frac{p}{2}$, resulting in the following Bloch Hamiltonian under the condition given by Eq.~(\ref{JzJxy}),
\begin{eqnarray}\label{Hk1kp}
\mathcal{H}_2(p)=E_F +J_{xy}\cos\frac{p}{2}\left(
      \begin{array}{cccccccc}
        \pm 4S &   -2\sqrt{S(2S-1)}  &   & &    &    &   &   \\
         -2 \sqrt{S(2S-1)}  & \pm(4S-1) & -2S &  &   &   &   &   \\
          &  -2S & \pm 4S &  -2S  & &   &   &   \\
          &   & -2S &  \pm 4S &  &    &   &   \\
           &   &    &  & \ddots&   &   &   \\
          &   &   &     & &\pm 4S &  -2S &   \\
          &   &   &    &  &-2S& \pm 4S &   -2\sqrt{2}S \\
          &   &   &    & &  &   -2\sqrt{2}S& \pm 4S \\
      \end{array}
    \right).
\end{eqnarray}
It is easy to check that
\begin{eqnarray}\label{PsiZES}
|\Psi^{(\pm)}_{\mathrm{ZES}}\rangle=2\sqrt{\frac{S}{2SN-1}} \left(\pm\frac{\sqrt{2S-1}}{2\sqrt{S}},1,\pm1,\cdots,\pm1,1,\pm\frac{\sqrt{2}}{2}\right)^T
\end{eqnarray}
\end{widetext}
gives two normalized zero-energy Bloch states satisfying $[\mathcal{H}_2(p(k))-E_F]|\Psi^{(\pm)}_{\mathrm{ZES}}\rangle=0$. The consistency between Eqs.~(\ref{Lk2a}), (\ref{Lk2b}), and (\ref{PsiZES}) indicates that Eq.~(\ref{pkn}) does hold for $n=2$.
\par For $n=3$, it is too tedious to write down the explicit expression for $(L_k)^3|F\rangle$. Nevertheless, we numerically confirm that the relation $p(k)=3k~(\mathrm{mod}~2\pi)$ holds, so that $p(k)\in K_2$. Explicitly
\begin{eqnarray}\label{pk3main}
  p=\begin{cases}
    3k, & \text{$k\in[-\frac{\pi}{3},0]$},\\
    3k+2\pi, & \text{$k\in[-\pi,-\frac{\pi}{3})$}.
  \end{cases}
\end{eqnarray}
For $k\in [-\pi,-\frac{2}{3}\pi]$ and $k\in[-\frac{\pi}{3},0]$, we find that the zero-energy state $(L_k)^3|F\rangle$ is also the ground state of $\mathcal{H}_3(p(k))-E_F$. However, for the middle region $k\in(-\frac{2}{3}\pi,-\frac{\pi}{3})$ the state $(L_k)^3|F\rangle$ is found to be an excited state. We believe some of these properties persist in the $n$-magnon sector with $n>3$. For example, the zero-energy state $(L_k)^n|F\rangle$ with $k\in[-\frac{\pi}{n},0]$ (suppose $N$ is divisible by $n$) should be the ground state of $\mathcal{H}_n(nk)-E_F$.
\section{Two-magnon sector}\label{SecIV}
\par In this section, we use our formalism to study the emergence of two-magnon bound states. We also calculate the transverse dynamic structure factor near the saturation magnetization in a higher-spin antiferromagnetic chain, where the ground state is approximated by the sub-ground state in the one-magnon sector.
\subsection{Two-magnon bound states}
\par In the limiting case of $J_{xy}=0$ and for $J_z,D>0$, all the $\mathcal{H}_2(k)$'s become diagonal and the Bloch states $|\psi_0(k)\rangle$'s and $|\psi_1(k)\rangle$'s form two $N$-fold degenerate manifolds with excitation energies $\Omega_0$ and $\Omega_1$, respectively. As can be seen from Fig.~\ref{Fig1D}, these states (the red and blue solid circles) correspond to edge states on the effective free-end lattice. When a finite but small $J_{xy}/J_z$ is introduced, two distinct types of two-magnon bound states, the so-called single-ion (corresponding to $r=0$) and the exchange (corresponding to $r=1$) bound states, will emerge, as revealed by different methods in previous studies~\cite{Silberglitt1970,Papan1987}. By noting that $\Omega_1-\Omega_0=2D-J_z$, the single-ion bound states should be dominated for $D/J_z>1/2$.
\begin{figure}
\includegraphics[width=.52\textwidth]{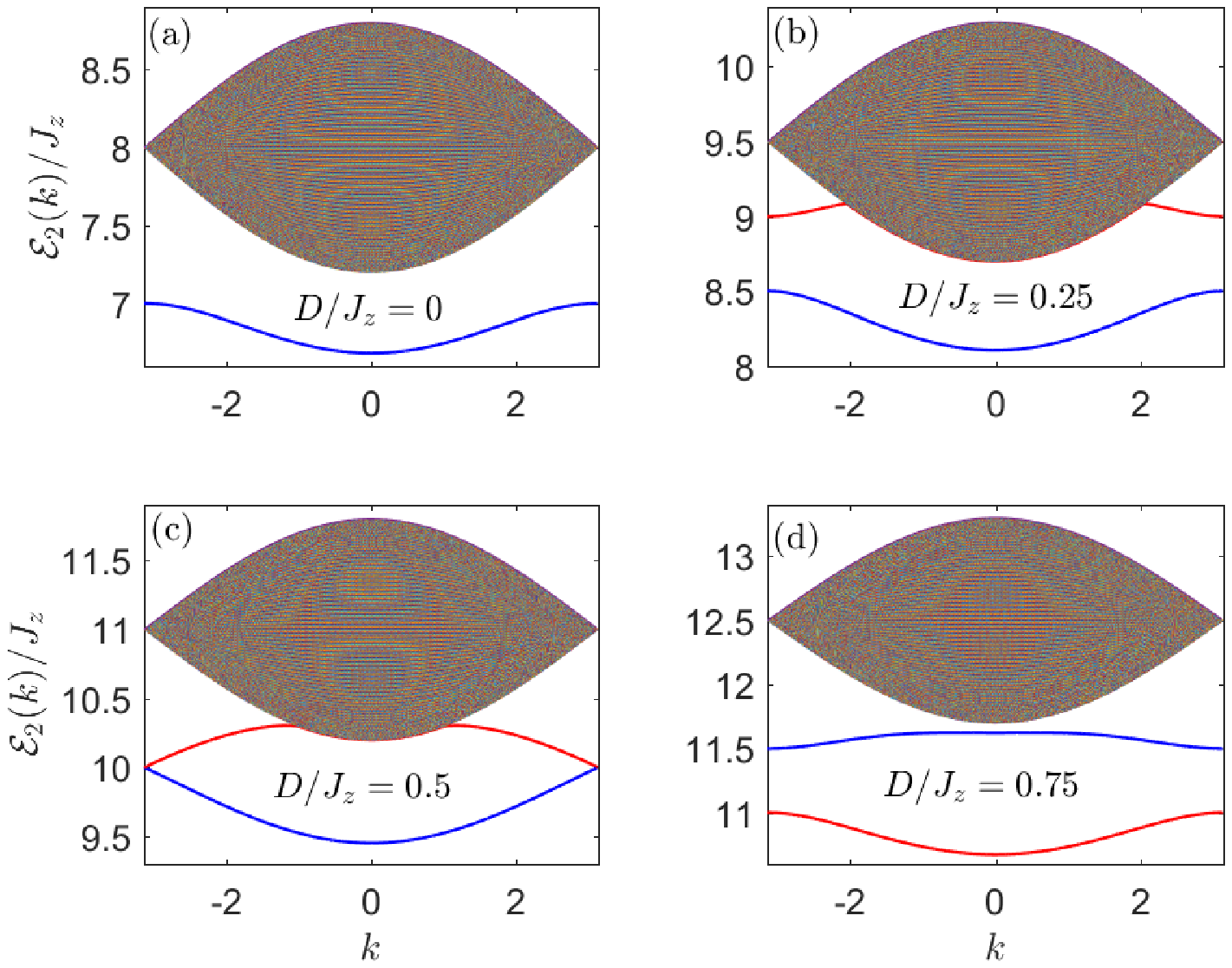}
\caption{Two-magnon excitation spectra for various values of $D/J_z$. The red and blue curves indicate the single-ion and exchange two-magnon bound states, respectively. Parameters: $N=1000$, $S=2$, and $J_{xy}/J_z=0.1$.}
\label{twomagnonspectra}
\end{figure}
\par Figure~\ref{twomagnonspectra} shows the two-magnon excitation spectra $\mathcal{E}_2(k)/J_z$ for $J_{xy}/J_z=0.1$ and several values of $D/J_z$. In the absence of the single-ion anisotropy [Fig.~\ref{twomagnonspectra}(a)], we only observe the exchange bound state $|\Psi_{\mathrm{exc}}(k)\rangle$ due to the large gap $\Omega_1-\Omega_0$. The corresponding  wave function $\langle\psi_r(k)|\Psi_{\mathrm{exc}}(k)\rangle$ (note that it is real) is localized around site $r=1$ on the effective lattice. Increasing $D/J_z$ to 0.25 decreases the gap $\Omega_1-\Omega_0$ and both the exchange bound state $|\Psi_{\mathrm{exc}}(k)\rangle$ and single-ion two-magnon bound state $|\Psi_{\mathrm{s}-\mathrm{ion}}(k)\rangle$ emerge at the edge of the Brillouin zone [Fig.~\ref{twomagnonspectra}(b)]. For $D/J_z=0.5$, we have $\Omega_1=\Omega_0$, so that the two separated branches of the spectra touch each other at $k=-\pi$ and  the two wave functions are approximately equally distributed between the two sites ($r=0$ and 1) within the zone [Fig.~\ref{twomagnonspectra}(c)]. When $D/J_z$ increases to 0.75, the lowest energy level is occupied by the single-ion bound states [Fig.~\ref{twomagnonspectra}(d)].
\subsection{Transverse dynamic structure factor near saturation magnetization}\label{SecIVB}
\par Recently, it was shown in Ref.~\cite{Balents} that two-magnon bound states in an antiferromagnetic spin-1/2 chain appear as a higher energy branch in the transverse dynamic structure factor. In this section, we use our formalism to calculate the transverse dynamic structure factor near saturation magnetization for an antiferromagnetic XXZ chain with higher spins. As we will see, the usual exchange and the single-ion two-magnon bound states appear, respectively, as high and low energy branches in the transverse dynamic structure factor.
\par To this end, we add a Zeeman term to the original Hamiltonian and allow for negative values of $J_{xy}$ and $J_z$,
\begin{eqnarray}
H\to H-B\sum^N_{j=1}S^z_j,
\end{eqnarray}
so that the one-magnon excitation energy becomes
\begin{eqnarray}
\mathcal{E}_1(k)=2S(J_z-J_{xy}\cos k)+D(2S-1)+B.
\end{eqnarray}
\begin{figure}
\includegraphics[width=.52\textwidth]{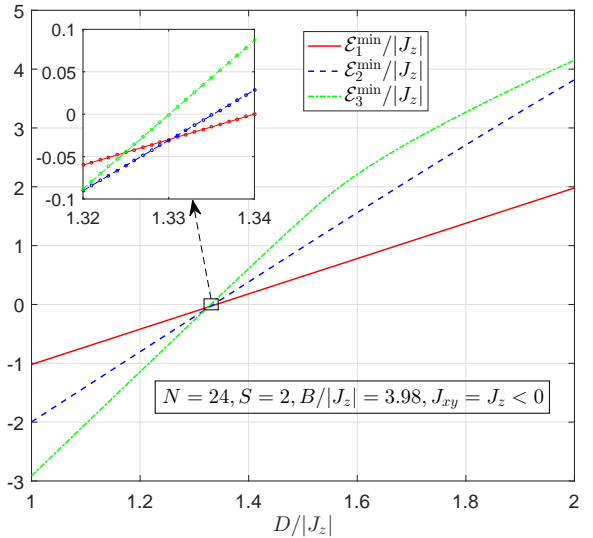}
\caption{Minimal excitation energies in the one- (solid red), two- (dashed blue), and three-magnon (dash-dotted green) sectors as functions of $D/|J_z|$. We considered an antiferromagnetic XXX chain with $J_{xy}=J_z<0$ and choose $B=3.98|J_z|$, which is just below the saturation magnetic field $B_{\mathrm{sat}}= 4S|J_z|$ at $D=0$. The inset shows the magnification of the crossover range around $D/|J_z|=1.33$. }
\label{minE123}
\end{figure}
\par For large enough $B$, the ground state is the polarized state $|F\rangle$. Below we consider an antiferromagnetic chain with $J_{xy}, J_z<0$, then the lowest-energy one-magnon state is achieved for $k=-\pi$. The minimal excitation energy $\mathcal{E}^{\min}_1=\mathcal{E}_1(-\pi)=2S(J_z+J_{xy})+D(2S-1)+B$ depends linearly on both $B$ and $D$. The saturation magnetic field is defined by
\begin{eqnarray}
B_{\mathrm{sat}}=2S|J_z+J_{xy}|-D(2S-1).
\end{eqnarray}
\par However, the minimal excitation energies in the two- and three-magnon sectors, $\mathcal{E}^{\min}_2$ and $\mathcal{E}^{\min}_3$, depend linearly only on $B$, see Fig.~\ref{minE123} for an illustration. For a magnetic field just below the saturation value $B_{\mathrm{sat}}=4S|J_z|$ (for $D=0$), we find that there exists a narrow range of $D/|J_z|\in (1.331,1.339)$ within which the one-magnon excitation energy is not only negative but also the smallest among $\{\mathcal{E}^{\min}_1,\mathcal{E}^{\min}_2,\mathcal{E}^{\min}_3\}$ (inset of Fig.~\ref{minE123}). This indicates that the lowest one-magnon state is the most energetically favorable in the above parameter range, which, however, becomes narrower as $N$ increases.
\par
\begin{figure*}
\includegraphics[width=1.035\textwidth]{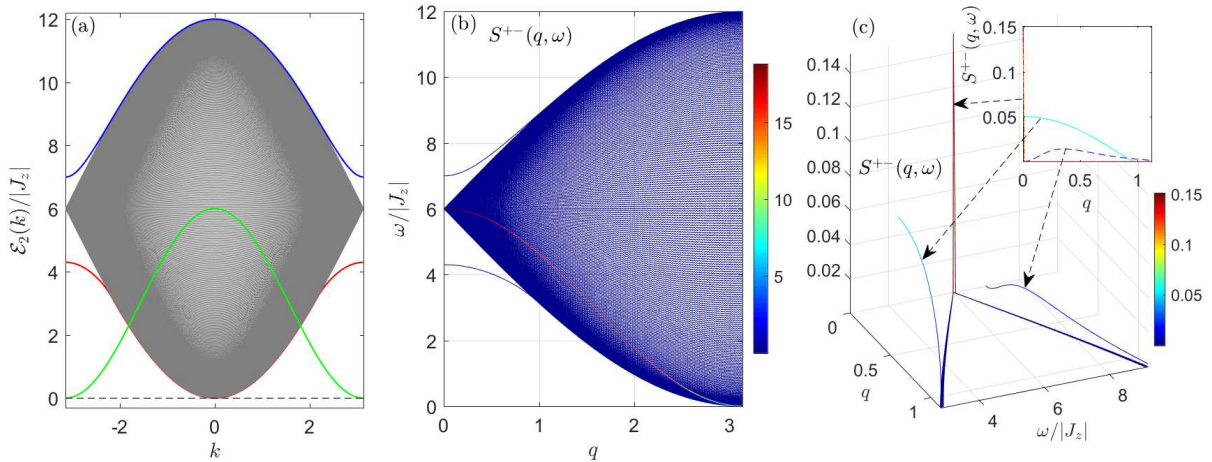}
\caption{(a) Two-magnon excitation spectra for a spin-$3/2$ XXX chain ($J_{xy}=J_z<0$) with $N=500$, $D/|J_z|=0.84999$, and $B/|J_z|=4.3$. The two separated branches correspond to the single-ion (lower red) and exchange (upper blue) bound states. The solid green and dashed horizontal lines represent the one-magnon excitation spectra and zero-excitation-energy point, respectively. The lowest one-magnon (two-magnon) excitation energy is $-2\times 10^{-5}|J_z|$ ($7.704\times 10^{-5}|J_z|$). (b) Dynamic structure factor $S^{+-}(q,\omega)$ calculated by Eq.~(\ref{Spmqw}) for the lowest one-magnon state $|\psi(-\pi)\rangle$. The highlighted curve shows the usual contribution~\cite{Balents}. (c) The three-dimensional plot of $S^{+-}(q,\omega)$ contributed by the two bound states and part of the continuum (on a different color scale). The inset shows that as $q$ increases, $S^{+-}(q,\omega)$ for the exchange bound state (dashed blue) shows a nonmonotonic behavior, while the one for the single-ion bound state (solid cyan) decreases monotonically.}
\label{DSFfig}
\end{figure*}

 As an example, in Fig.~\ref{DSFfig}(a) we plot both the one- and two-magnon excitation spectra for a spin-$3/2$ antiferromagnetic XXX chain ($J_{xy}=J_z<0$) with $N=500$ sites. The parameter region within which $\mathcal{E}^{\min}_1<\min(0,\mathcal{E}^{\min}_2)$ is fulfilled becomes so narrow that we have to finely tune the value of $D/|J_z|$ for a fixed $B/|J_z|$. In our example, we set $D/|J_z|=0.84999$ and $B/|J_z|=4.3$, yielding $(\mathcal{E}^{\min}_1,\mathcal{E}^{\min}_2)/|J_z|=(-2\times 10^{-5},7.704\times 10^{-5})$ (we have checked that they are not numerical errors). It is easy to see that the upper (lower) separated branch corresponds to the exchange (single-ion) bound state since $\Omega_0\approx 4.3|J_z|$ and $\Omega_1\approx 7|J_z|$.
\par For a general eigenstate $|\Phi\rangle$ of $H$ with eigenenergy $E_{\Phi}$, the transverse dynamic structure factor is defined as~\cite{Balents}
\begin{eqnarray}\label{DSF}
 S^{+-}(q,\omega) &\equiv&\frac{1}{N}\int^\infty_{-\infty}dt e^{i(\omega+E_{\Phi}) t} \langle \Phi|  L^\dag_q e^{-iHt}L_q |\Phi\rangle\nonumber\\
 &=&\frac{2\pi}{N}\sum_E\delta(\omega+E_{\Phi}-E) |\langle E|L_q |\Phi\rangle|^2,
\end{eqnarray}
where $L_q=\sum^N_{j=1}e^{iq j}S^-_j$ and $H|E\rangle=E|E\rangle$.
\par We now assume that the ground state is well approximated by the one-magnon eigenstate $|\Phi\rangle=|\psi(Q)\rangle$ given by Eq.~(\ref{psik1})~\cite{Balents}. Since the operator $L_q$ carries momentum $q$, only two-magnon states with momentum $q+Q$ contribute to the summation over the eigenstates $\{|E\rangle\}$ in Eq.~(\ref{DSF}). If we let $|\Psi_{2,\alpha}(k)\rangle$ ($\alpha=1,2,\cdots,N/2$ for $k\in K'_1$ and $\alpha=1,2,\cdots,N/2+1$ for $k\in K_1$) be the eigenstate of the Bloch Hamiltonian $\mathcal{H}_2(k)-E_F$ with eigenenergy $\mathcal{E}_{2,\alpha}(k)$, a direct calculation leads to (for $q+Q\in K_1$)
\begin{widetext}
\begin{eqnarray}\label{Spmqw}
S^{+-}(q,\omega)&=& \frac{2\pi}{N}   \sum_\alpha \delta[\omega+ \mathcal{E}_1(Q) -\mathcal{E}_{2,\alpha}(Q+q)]\times\bigg| \sqrt{2(2S-1)}\langle \psi_0(Q+q)|\Psi_{2,\alpha}(Q+q)  \rangle  \nonumber\\
&&+2\sqrt{S}  e^{-i(Q-q)\frac{N}{4}}\langle \psi_{\frac{N}{2}}(Q+q) |\Psi_{2,\alpha}(Q+q)\rangle  +2\sqrt{2S}\sum_{0<r<\frac{N}{2}}\cos\frac{(Q-q)r}{2}  \langle\psi_{r} (Q+q)|\Psi_{2,\alpha}(Q+q) \rangle \bigg|^2.\nonumber\\
\end{eqnarray}
\end{widetext}
A similar expression holds for $Q+q\in K'_1$ (with the term $2\sqrt{S}  e^{-i(Q-q)\frac{N}{4}}\langle \psi_{\frac{N}{2}}(Q+q) |\Psi_{2,\alpha}(Q+q)\rangle$ being removed).
\par Figure~\ref{DSFfig}(b) shows the dynamic structure factor $S^{+-}(q,\omega)$ calculated by the above equation using $Q=-\pi$. The dominant branch is the usual contribution~\cite{Balents}. To see the contribution of the two bound states, we plot in Fig.~\ref{DSFfig}(c) a three-dimensional plot of the $S^{+-}(q,\omega)$ near the edges of the band on a different color scale. For the exchange bound states (rightmost curve), we observe a finite $S^{+-}(q,\omega)$ exhibiting a nonmonotonic behavior as $q$ increases. This is similar to the case of a spin-$1/2$ XXX chain at high magnetization~\cite{Balents}. Interestingly, we also observe a slightly larger contribution from the single-ion bound states (leftmost curve), which shows a monotonic decay with increasing $q$. These behaviors can be more clearly seen from the inset of Fig.~\ref{DSFfig}(c), where we plot $S^{+-}(q,\omega)$ as a function of $q$. It is thus possible to uncover the appearance of both types of two-magnon bound states from investigating the experimentally relevant transverse dynamic structure factor $S^{+-}(q,\omega)$.
\section{Three-magnon sector}\label{SecV}
\par We now turn to study the three-magnon sector in detail with the help of the Bloch Hamiltonians shown in Fig.~\ref{3magnon2D}. In this section, we will set $J_z,J_{xy}>0$.
\subsection{Three-magnon bound states}
\par The structures of three-magnon bound states are much richer than the two-magnon ones due to the variety of the on-site energies, as can be seen from Fig.~\ref{3magnon2D}. It is intuitive to first look at the case of vanishing $J_{xy}$ for which all the hoppings in Figs.~\ref{3magnon2D}(a) and \ref{3magnon2D}(b) are turned off. According to the level diagram shown in Fig.~\ref{3magnon2D}(c), the ground state of $\mathcal{H}_3(k)|_{J_{xy}=0}$ for $D/J_z>1/2$ ($D/J_z<1/2$) is $|\psi_{0,0}(k)\rangle$ (are $|\psi_{0,1}(k)\rangle$ and $|\psi_{0,N-1}(k)\rangle$), which will be referred to as the  $p$ state (purple) [$b$ states (blue)] according to the colors of the circles.
\par
\begin{figure}
\includegraphics[width=.53\textwidth]{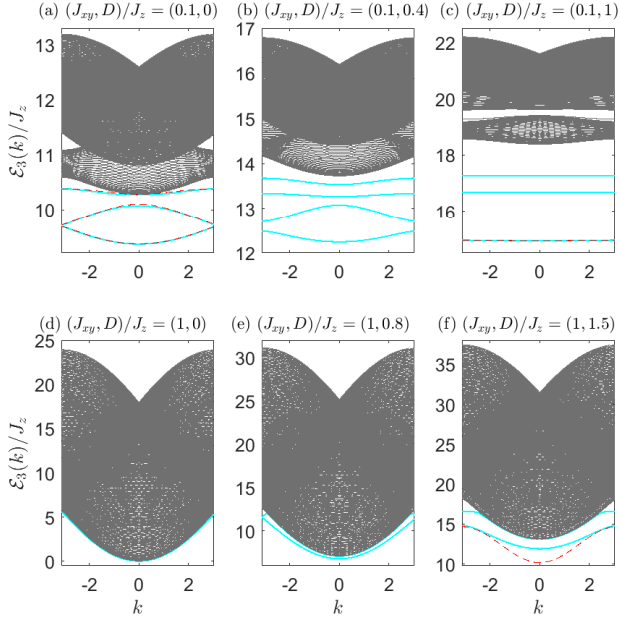}
\caption{Three-magnon excitation spectra for $N=60$ and $S=2$. The separated branches corresponding to three-magnon bound states are highlighted in cyan. The red dashed curves in (a) show the three eigenenergies of the effective Hamiltonian $\mathcal{H}^{(\mathrm{eff})}_{3,D=0}(k)-E_F$ given by Eq.~(\ref{Heff}), and the ones in (c) and (f) show the approximate energy given by Eq.~(\ref{E4th}).}
\label{Fig3}
\end{figure}

 Turning on the $xy$ coupling generally mixes these states to form quasi-continuous bands. However, bound states separated from the continua can emerge in certain parameter regimes. Figure~\ref{Fig3} shows the three-magnon excitation spectra calculated by diagonalizing the Bloch Hamiltonians $\{\mathcal{H}_3(k)\}$ for $N=60$ and $S=2$. We observe several separated branches that indicate the emergence of three-magnon bound states.
\par Let us first discuss the case of vanishing $D$ [Fig.~\ref{Fig3}(a)]. For $J_{xy}/J_z=0$, the two $b$-states are degenerate with the $g$ state [green, $|\psi_{1,1}(k)\rangle$, see Fig.~~\ref{3magnon2D}(c)]. To analyze the properties of the system for small $J_{xy}/J_z$, we need to resort to degenerate perturbation theory. Here, we employ Takahashi's many-body perturbation theory~\cite{Taka} to derive an effective Bloch Hamiltonian $\mathcal{H}^{(\mathrm{eff})}_{3,D=0}(k)-E_F$ up to the third order in $J_{xy}/J_z$ in this three-dimensional degenerate manifold. Explicitly, consider a generic Hamiltonian $h=h_0+\lambda V$, where $\lambda V$ can be viewed as a perturbation. Let $P_0$ be the projector on the degenerate manifold associated with eigenvalue $E_0$ of $h_0$, then the Takahashi effective Hamiltonian up to the third order in $\lambda$ reads~\cite{Taka}
\begin{eqnarray}\label{heff}
h_{\mathrm{eff}}&=&E_0P_0+\lambda P_0VP_0+\lambda^2 P_0VS^1VP_0\nonumber\\
&&+\lambda^3\big(P_0VS^1VS^1VP_0-\frac{1}{2}P_0VS^2VP_0VP_0\nonumber\\
&&-\frac{1}{2}P_0VP_0VS^2VP_0\big),
\end{eqnarray}
where $S^k=\left(\frac{1-P_0}{E_0-h_0}\right)^k,~k\geq 1$.
\par We now apply the above theory to the Bloch Hamiltonian $\mathcal{H}_3(k)-E_F$ in the case of $D=0$. The nonperturbative ground-state manifold is spanned by $\{|\psi_{0,1}\rangle,|\psi_{0,N-1}\rangle,|\psi_{1,1}\rangle\}$ with a common energy $E_0=(6S-2)J_z$, so that $P_0=|\psi_{0,1}\rangle\langle\psi_{0,1}|+|\psi_{0,N-1}\rangle\langle\psi_{0,N-1}|+|\psi_{1,1}\rangle\langle\psi_{1,1}|$. After a straightforward calculation we obtain a $3\times3$ effective Hamiltonian $\mathcal{H}^{(\mathrm{eff})}_{3,D=0}(k)-E_F$ (see Appendix~\ref{AppB} for its explicit form).
\par The red dashed curves in Fig.~\ref{Fig3}(a) represent the three eigenenergies of $\mathcal{H}^{(\mathrm{eff})}_{3,D=0}(k)-E_F$ for $J_{xy}/J_z=0.1$, which are in good agreement with the exact results.
As $D/J_z$ increases to $0.4$ [Fig.~\ref{Fig3}(b)], the $p$, $g$, and $b$ states are responsible for the four separated levels. In the large anisotropy limit with $D/J_z=1$ [Fig.~\ref{Fig3}(c)], the lowest branch of the spectrum is dominated by the nondegenerate $p$ state. Using standard nondegenerate perturbation theory, we derive the ground-state energy correction up to the fourth order in $J_{xy}/(J_z-2D)$,
\begin{eqnarray}\label{E4th}
 \mathcal{E}_3(k)& \approx&  6SJ_z+D(6S-9)+\frac{3S(S-1)J^2_{xy}}{J_z-2D}\nonumber\\
& -&\frac{ 3S(S-1)(2S-1)  J_{xy}^3\cos k}{2(J_z-2D)^2}+\frac{3S(S-1)J^4_{xy}}{4( J_z-2D)^2}\times\nonumber\\
 &&\left[\frac{2S(2S-1)}{ J_z-3D}-\frac{S^2}{2D}-\frac{2S^2-2S-1}{ J_z-2D}\right].
\end{eqnarray}
Note that the dispersion arises from the third order, and the second- and fourth- order corrections only give an energy shift. The red dashed curve in Fig.~\ref{Fig3}(c) shows the result given by Eq.~(\ref{E4th}), which agrees well with the exact result. The middle quasi-continuous band around $\mathcal{E}_3(k)/J_z=19$ is due to the mixing of the $N-3$ edge $r$ states (red) and the $g$ state.
\par The lower panels of Fig.~\ref{Fig3} show the spectrum for $J_{xy}/J_z=1$. Compared with the case of small $J_{xy}/J_z$, a larger $D$ is needed to observe the bound states. Nevertheless, the lowest branches in Figs.~\ref{Fig3}(e) and \ref{Fig3}(f) are still dominated by the $p$ states. The fourth-order perturbation still gives accurate results for the spectrum at the edges of the momentum space [red dashed curve in Fig.~\ref{Fig3}(f)].
\subsection{Transverse dynamic structure factor for the two-magnon states}
\par In Sec.~\ref{SecIVB} we calculated the transverse dynamic structure factor for the lowest one-magnon state $|\psi(-\pi)\rangle$ using the Bloch states. In this subsection we will calculate the transverse dynamic structure factor for a state in the two-magnon sector. We assume that the two-magnon state of interest is some eigenstate $|\Psi_{2,\xi}(Q)\rangle$ of $\mathcal{H}_2(Q)-E_F$ with excitation energy $\mathcal{E}_{2,\xi}(Q)$, where $\xi$ labels this particular eigenstate and we take it as the lowest one in the $Q$-subspace. By setting $|\Phi\rangle=|\Psi_{2,\xi}(Q)\rangle$ in Eq.~(\ref{DSF}), we have
\begin{eqnarray}
S^{+-}(q,\omega)&=&\frac{2\pi}{N} \sum_\alpha\delta[\omega+\mathcal{E}_{2,\xi}(Q)-\mathcal{E}_{3,\alpha}(q+Q)]\nonumber\\
&&\times|\langle\Psi_{3,\alpha}(q+Q)| L_q|\Psi_{2,\xi}(Q)\rangle|^2,
\end{eqnarray}
where $|\Psi_{3,\alpha}(q+Q)\rangle$ is the eigenstate of $\mathcal{H}_3(q+Q)-E_F$ with eigenenergy $\mathcal{E}_{3,\alpha}(q+Q)$. The explicit form of the matrix element $\langle\Psi_{3,\alpha}(q+Q)| L_q|\Psi_{2,\xi}(Q)\rangle$ is lengthy and not illuminating, but can be easily handled in the numerical simulation.
\par To be specific, we consider a ferromagnetic $S=3/2$ XXZ chain with $J_z>0$ and $N=90$. The parameters are chosen as $J_{xy}/J_z=0.5$, $D/J_z=1.5$, and $B/J_z=1$, for which the lowest two-magnon eigenstate lies in the $Q=0$ subspace, giving $\mathcal{E}_{2,\xi}(0)/J_z=9.8209$ and $|\Psi_{2,\xi}(0)\rangle=\sum^{N/2}_{r=0}c_{r}|\psi_r(0)\rangle$ with $c_0=-0.7995,~c_1=-0.5443,~c_2=-0.2303,~c_3=-0.0975,~\cdots$. Here, we only show the first few components since the amplitude of $|\psi_r(0)\rangle$ decays rapidly as $r$ increases. As a good approximation, we take into account only the amplitudes up to $r=2$, yielding
\begin{widetext}
\begin{eqnarray}\label{ME32}
\langle\Psi_{3,\alpha}(q)| L_q|\Psi_{2,\xi}(0)\rangle&=&c_0 \sqrt{6(S-1)}\langle\Psi_{3,\alpha}(q) |\psi_{0,0}(q)\rangle  \nonumber\\
&&+\sum_{r=1,2}   \left[\sqrt{2S}c_0e^{i\frac{2}{3}rq}+\sqrt{2(2S-1)}c_r e^{-i\frac{r}{3}q}\right]\langle\Psi_{3,\alpha}(q)|\psi_{0,r}(q)\rangle \nonumber\\
&&+\sum_{r=1,2} \left[\sqrt{2S}c_0e^{i\frac{2}{3}(N-r)q}+\sqrt{2(2S-1)}c_r e^{-i\frac{N-r}{3}q}\right] \langle\Psi_{3,\alpha}(q)|\psi_{0,N-r}(q)\rangle\nonumber\\
&&+c_0  \sqrt{2S} \sum^{N-3}_{n=3} e^{i\frac{2}{3}nq } \langle\Psi_{3,\alpha}(q) |\psi_{0,n}(q)\rangle  \nonumber\\
&&+\sqrt{2S}\sum_{r=1,2} c_r\sum^{N-2r-1}_{l=r+1}e^{\frac{i}{3}(r+2l)q}\langle\Psi_{3,\alpha}(q)|\psi_{r,l}(q)\rangle+2\sqrt{2S} \sum_{r=1,2}c_r \cos qr\langle\Psi_{3,\alpha}(q) |\psi_{r,r}(q)\rangle\nonumber\\
&&+c_2\sqrt{2S}\langle\Psi_{3,\alpha}(q) |[|\psi_{1,1}(q)\rangle+ e^{-i\frac{N-4}{3}q}|\psi_{1,N-3}(q)\rangle+ e^{-i\frac{4}{3}q}|\psi_{1,2}(q)\rangle]+\cdots
\end{eqnarray}
\end{widetext}
There also exist two branches of low-lying three-magnon bound states, $|\Psi_{3,1}(q)\rangle$ and $|\Psi_{3,2}(q)\rangle$, which are dominated by the $p$ states (around $\mathcal{E}_{3}(k)/J_z\sim 11.5$) and the $b$ states (around $\mathcal{E}_{3}(k)/J_z\sim 14.5$), respectively. For these two branches, the matrix element given by Eq.~(\ref{ME32}) is mainly contributed by the terms proportional to $\langle\Psi_{3,1}(q)|\psi_{0,0}(q)\rangle$, $\langle\Psi_{3,2}(q)|\psi_{0,1}(q)\rangle$, and $\langle\Psi_{3,2}(q)|\psi_{0,N-1}(q)\rangle$ [the first three lines of Eq.~(\ref{ME32})], giving rise to the evidence of three-magnon bound states in $S^{+-}(q,\omega)$.
\begin{figure}
\includegraphics[width=.51\textwidth]{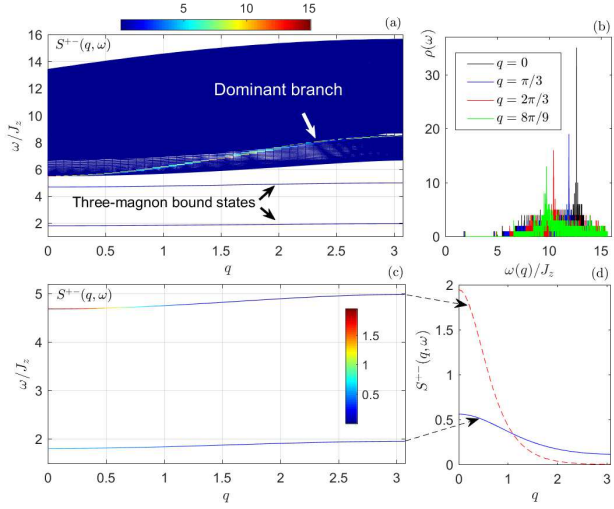}
\caption{(a) Transverse dynamic structure factor $S^{+-}(q,\omega)$ for the lowest two-magnon state $|\psi_{2,\xi}(0)\rangle$ for a spin-3/2 ferromagnetic XXZ chain. (b) Density of states $\rho(\omega)$ for the $\omega(q)$ shown in (a). (c) The two separated branches contributed by the two three-magnon bound states. The inset shows that $S^{+-}(q,\omega)$ is significantly different from zero for small $q$, indicating the presence of three-magnon bound states. Parameters: $N=90$, $J_{xy}/J_z=0.5$, $D/J_z=1.5$, and $B/J_z=1$.}
\label{DSF3}
\end{figure}
\par Figure~\ref{DSF3}(a) shows the dynamic structure factor calculated by using the full expression of $\langle\Psi_{3,\alpha}(q)| L_q|\Psi_{2,\xi}(0)\rangle$. We observe a dominant single branch within the continuum, along with two separated branches (with smaller magnitudes) arising from the two types of three-magnon bound states. To see how the three-magnon states contribute to the dominant branch, we plot in Fig.~\ref{DSF3}(b) the numerical density of states $\rho(\omega)$ for several $q$'s (we choose the frequency interval as $\Delta\omega(q)=[\omega_{\max}(q)-\omega_{\min}(q)]/2000$ and count the number of states in each interval). It can be seen that the peak of $\rho(\omega(q))$ moves to the low-energy regime as $q$ increases, indicating that the dominant branch in $S^{+-}(q,\omega)$ seems irrelevant to the density of states.
\par A detailed numerical analysis shows that the dominant branch is mainly contributed by the fourth and fifth lines of Eq.~(\ref{ME32}), where a constructive interference of the amplitudes $\langle\Psi_{3,\alpha}(q)|\psi_{r_1,r_2}(q)\rangle$ occurs. Actually, the fourth line of Eq.~(\ref{ME32}) is mainly contributed by three-magnon eigenstates that have significant overlap with the Bloch states $\{|\psi_{0,n}(q)\}$ with $n=3,\cdots,N-3$ [red circles in Fig.~\ref{3magnon2D}(a)]. These eigenstates therefore can be thought of as a mixture of a single-ion two-magnon bound state and a one-magnon state. Similarly, the terms with $r=1$ in the fifth line of Eq.~(\ref{ME32}) are contributed by eigenstates that are mixtures of an exchange two-magnon bound state and a single magnon [orange circles in Fig.~\ref{3magnon2D}(b)]. Of course, there are also partial contributions from the three-magnon scattering states to the matrix element.
\par The magnitudes of the two lower energy branches in Fig.~\ref{DSF3}(a) are much smaller than the dominant branch. However, they are much larger than $S^{+-}(q,\omega)$ in the continuous region with the dominant branch excluded. The two separated branches are shown in Fig.~\ref{DSF3}(c) on a different color scale. From Fig.~\ref{DSF3}(d) we see that $S^{+-}(q,\omega)$ is significantly different from zero for small $q$. These results indicate that signatures of the three-magnon bound states can also be detected in the transverse dynamic structure factor.
\subsection{Three-magnon quantum walks}
\par Besides calculating the transverse dynamic structure factor, our formalism also allows us to evaluate the real-time dynamics of local spin excitations by simulating independent quantum walks on the effective lattices. The foregoing identification of magnon bound states provides an intuitive way to look at the multimagnon dynamics. Suppose the system is initially prepared in a general localized state with $(r_1,r_2)\neq(m,m)$,
\begin{eqnarray}
|\Phi(0)\rangle=|\phi^{n}_{r_1,r_2}\rangle.
\end{eqnarray}
We are interested in the local magnetization dynamics $\langle S^z_j(t)\rangle=\langle\Phi(0)|e^{iHt}S^z_je^{-iHt}|\Phi(0)\rangle$. By expanding $|\phi^{n}_{r_1,r_2}\rangle$ in terms of the Bloch states using Eq.~(\ref{phinrrexp}), we are able to derive the following expression for $\langle S^z_j(t)\rangle$,
\begin{eqnarray}
\langle S^z_j(t)\rangle&=&S- \sum^3_{a=1}\sum^{(m-1,m+1)}_{s_1,s_2=(0,0)}|X^{(a),j,n}_{r_1r_2;s_1s_2}(t)|^2-|Y^{j,n}_{r_1r_2}(t)|^2.\nonumber\\
\end{eqnarray}
Here,
\begin{eqnarray}
X^{(1),j,n}_{r_1r_2;s_1s_2}(t)&\equiv& \frac{1}{N}\sum_{k\in K_0} e^{ i\frac{k}{3}[3(j-n-1)+2s_1+s_2-r_1-r_2]}\nonumber\\
&&F_{r_1r_2;s_1s_2}(k,t),\nonumber\\
X^{(2),j,n}_{r_1r_2;s_1s_2}(t)&\equiv& \frac{1}{N}\sum_{k\in K_0}  e^{i\frac{k}{3}[3(j-n-1)-s_1+s_2-2 r_1  -r_2 ]}\nonumber\\
&&F_{r_1r_2;s_1s_2}(k,t),\nonumber\\
X^{(3),j,n}_{r_1r_2;s_1s_2}(t)&\equiv& \frac{1}{N}\sum_{k\in K_0}  e^{i\frac{k}{3}[3(j-n-1)-s_1-2s_2-2r_1 -r_2 ]}\nonumber\\
&&F_{r_1r_2;s_1s_2}(k,t),
\end{eqnarray}
and
\begin{eqnarray}
Y^{j,n}_{r_1r_2}(t)&\equiv&\frac{1}{\sqrt{Nm}} \sum_{k\in K_2} e^{ik(j-n-1)}e^{i\frac{k}{3}(3m-2r_1-r_2)}\nonumber\\
&&F_{r_1r_2;mm}(k,t),
\end{eqnarray}
with
\begin{eqnarray}\label{Fkt}
F_{r_1r_2;s_1s_2}(k,t)&\equiv&\sum_{\alpha}e^{-iE_{3,\alpha}(k) t} \langle\Psi_{3,\alpha}(k)|\psi_{r_1,r_2}(k)\rangle\nonumber\\
&&\langle\psi_{s_1s_2}(k)|\Psi_{3,\alpha}(k)\rangle.
\end{eqnarray}
\begin{figure}
\includegraphics[width=.50\textwidth]{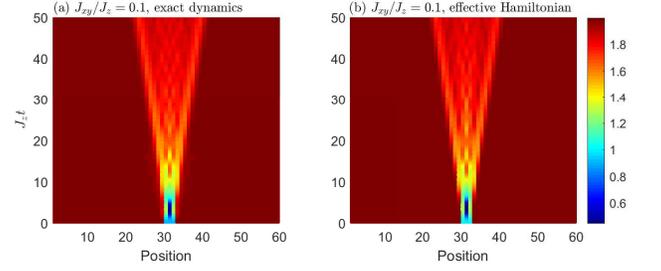}
\caption{(a) Magnetization dynamics $\langle S^z_j(t)\rangle$ from an initial state $|\phi^{\frac{N}{2}-1}_{1,1}\rangle=|\frac{N}{2},\frac{N}{2}+1,\frac{N}{2}+2\rangle$ for $N=60$ and $J_{xy}/J_z=0.1$. (b) The corresponding approximated dynamics using the effective Hamiltonian $\mathcal{H}^{(\mathrm{eff})}_{3,D=0}(k)$.}
\label{Figeffdynamics}
\end{figure}
The initial condition for $F_{r_1r_2;s_1s_2}(k,t)$ is given by
\begin{eqnarray}
F_{r_1r_2;s_1s_2}(k,t=0)&=&\delta_{r_1s_1}\delta_{r_2s_2}.
\end{eqnarray}
We see that the $F$'s given by Eq.~(\ref{Fkt}) are mainly contributed by eigenstates having significant overlap with the initial component state $|\psi_{r_1,r_2}(k)\rangle$. In particular, if the initial state is some real-space bound state, e.g., the local state $|\frac{N}{2},\frac{N}{2}+1,\frac{N}{2}+2\rangle$ with three successive spin excitations, it is then reasonable to expect that the corresponding three-magnon bound states (the eigenstates) will mainly contribute to the magnetization dynamics, provided these bound states are well separated from the continuum.
\par Figure \ref{Figeffdynamics}(a) shows the evolution of $\langle S^z_j(t)\rangle$ starting with $|\Phi(0)\rangle= |\frac{N}{2},\frac{N}{2}+1,\frac{N}{2}+2\rangle$ for $D=0$ and $J_{xy}/J_z=0.1$ [corresponding to Fig.~\ref{Fig3}(a)]. The situation here is similar to a three-boson quantum walk recently studied in Ref.~\cite{Guan2021}. We expect that the three-magnon bound states shown in Fig.~\ref{Fig3}(a) can accurately capture the magnetization dynamics since $|\Phi(0)\rangle$ is a linear combination of the $g$-states. To this end, we use the $3\times 3$ effective Hamiltonian $\mathcal{H}^{(\mathrm{eff})}_{3,D=0}(k)$ given by Eq.~(\ref{Heff}) to approximately calculate $\langle S^z_j(t)\rangle$ [Fig.~\ref{Figeffdynamics}(b)], which is found to agree well with the result obtained by full quantum simulation. However, deviation from the exact dynamics is observed for a larger $J_{xy}/J_z$, due to the fact that the highest effective level starts merging into the continuous band (data not shown).
\begin{figure}
\includegraphics[width=.52\textwidth]{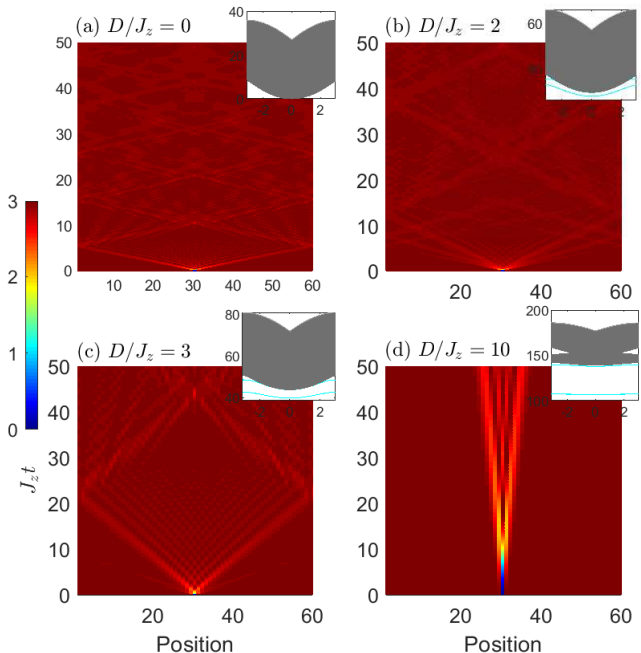}
\caption{Magnetization dynamics $\langle S^z_j(t)\rangle$ from an initial state $|\frac{N}{2},\frac{N}{2},\frac{N}{2}\rangle$ for $J_{xy}/J_z=1$, $N=60$, and $S=3$. (a) $D/J_z=0$, (b) $D/J_z=2$, (c) $D/J_z=3$, (d) $D/J_z=10$. The corresponding excitation spectra as a function of the wave number $k$ are shown in the upper-right corner of each panel.}
\label{Fig4}
\end{figure}
\par Figure \ref{Fig4} shows $\langle S^z_j(t)\rangle$ starting with $|\Phi(0)\rangle=|\phi^{\frac{N}{2}-1}_{0,0}\rangle=|\frac{N}{2},\frac{N}{2},\frac{N}{2}\rangle$ for $J_{xy}/J_z=1$, $S=3$ and several values of $D/J_z$. In the upper-right corner of each panel we also plot the corresponding three-magnon excitation spectrum. It can be seen that the propagation of the magnetization profile narrows down as $D/J_z$ increases. In the absence of the single-ion anisotropy, the large XX interaction destroys the formation of bound states, so that the dynamics is mainly contributed by the scattering states [Fig.~\ref{Fig4}(a)]. For $D/J_z=2$, we observe two new wave fronts due to the appearance of the two bound states [Fig.~\ref{Fig4}(b)]. The dynamics in the case of $D/J_z=3$ behaves similarly but with a slower propagating velocity due to the slightly flattened dispersion~[Fig.~\ref{Fig4}(c)]. In the large-$D$ limit, both the continuum band and the bound states dispersions become nearly flat, leading to a confined propagation around the center of the chain [Fig.~\ref{Fig4}(d)].
\par To understand the short-time dynamics in the small and large $D$ limits, we perform a time-dependent perturbative analysis, which gives the following initial Gaussian evolution,
\begin{eqnarray}
\langle S^z_{\frac{N}{2}}(t)\rangle &\approx&S-1-2e^{-(J_{xy}t_g)^2t^2},\nonumber\\
\langle S^z_{\frac{N}{2}\pm 1}(t)\rangle &\approx&S-1+e^{-(J_{xy}t_g)^2t^2}.
\end{eqnarray}
The spin flips therefore mainly spread to nearest neighbors at short times.
\section{Conclusions and discussions}\label{SecVI}
\par In this work, we provide the construction of exact Bloch states for the three-magnon sector in a finite-size higher-spin periodic XXZ chain. Each Bloch Hamiltonian defines a single-particle problem on a triangle-shape lattice. Several types of magnon bound states are identified as edge states on the lattice. We reveal the condition under which zero-energy states upon the ferromagnetic state emerge. The two-magnon sector is also studied using similar ideas. By computing the transverse dynamic structure factor, we find signatures of the multimagnon bound states for a chain with higher spins. With the help of our formalism, we also calculate the three-magnon dynamics by simulating single-particle quantum walks on the effective lattices. The spread of local spin excitations over the chain is explained in terms of propagations of three-magnon bound states in certain parameter regimes.
\par We finally mention some possible applications of our exact formalism. Our method can be directly applied to higher-spin chains with higher order terms or next-nearest-neighbor couplings, which provides an opportunity to rigorously study multimagnon bound states in finite-size frustrated ferromagnetic chains. It is also straightforward to extend our formalism to more general translationally invariant systems, such as itinerant particle systems described by the Fermi- or Bose-Hubbard models.

\noindent{\bf Acknowledgements:}
This work was supported by the Natural Science Foundation of China (NSFC) under Grant No. 11705007, and partially by the Beijing Institute of Technology Research Fund Program for Young Scholars. H.K. was supported in part by JSPS Grant in-Aid for Scientific Research on Innovative Areas No. JP20H04630, JSPSKAKENHI Grant No. JP18K03445, Grant-in-Aid for Transformative Research Areas (A) ``Extreme Universe" No. JP21H05191[D02], and the Inamori Foundation. X.-W.G. was supported by the NSFC Key Grant  No. 12134015 and the NSFC Grant No. 11874393.

\appendix
\begin{widetext}
\section{Proof of Eq.~~(\ref{ZEST})}\label{AppA}
\par We start with calculating the commutator
\begin{eqnarray}
[H,L_k]&=&(J_z e^{ik}-J_{xy})\sum^{N}_{n=1}e^{ikn}S^-_{n+1}S^z_n+(J_z-J_{xy}e^{ik})\sum^{N}_{n=1}e^{ikn} S^-_n S^z_{n+1}.
\end{eqnarray}
By applying $[H,L_k]$ to $|F\rangle$, we obtain
\begin{eqnarray}
[H,L_k]|F\rangle&=&S(J_z e^{ik}-J_{xy})\sum^{N}_{n=1}e^{ikn}S^-_{n+1}|F\rangle+S(J_z-J_{xy}e^{ik})\sum^{N}_{n=1}e^{ikn} S^-_n |F\rangle\nonumber\\
&=& 2S(J_z-J_{xy}\cos k)L_k|F\rangle=0,
\end{eqnarray}
which proves Eq.~(\ref{ZEST}) for $n=1$. We now observe that the commutator $[L_k,[H,L_k]]= -2 e^{ik}(J_z-J_{xy}\cos k)\sum_{ n}e^{i2kn} S^-_{n+1} S^-_n$ vanishes under the condition given by (\ref{JzJxy}), so that
\begin{eqnarray}
0&=&[L_k,[H,L_k]]|F\rangle=L_kHL_k|F\rangle-L_k^2H|F\rangle-HL^2_k|F\rangle+L_kHL_k|F\rangle=-HL^2_k|F\rangle+E_FL^2_k|F\rangle,
\end{eqnarray}
which proves Eq.~(\ref{ZEST}) for $n=2$. Following Refs.~\cite{Batista,Wouters}, we assume Eq.~(\ref{ZEST}) holds for $l$ and $l+1$, i.e., $H(L_k)^l|F\rangle=E_F(L_k)^l|F\rangle$ and $H(L_k)^{l+1}|F\rangle=E_F(L_k)^{l+1}|F\rangle$. Then,
\begin{eqnarray}
0&=&[L_k,[H,L_k]](L_k)^l|F\rangle=-HL^2_k(L_k)^l|F\rangle+L_kHL_k(L_k)^l|F\rangle=-(H-E_F)(L_k)^{l+2}|F\rangle.
\end{eqnarray}
By mathematical induction, we therefore proved Eq.~(\ref{ZEST}) for all $n\leq 2NS+1$.
\section{Explicit form of the effective Hamiltonian $\mathcal{H}^{(\mathrm{eff})}_{3,D=0}(k)-E_F$}\label{AppB}
The $3\times 3$ effective Hamiltonian $\mathcal{H}^{(\mathrm{eff})}_{3,D=0}(k)-E_F$ can be directly obtained by using Eq.~(\ref{heff}):
\begin{eqnarray}\label{Heff}
~[\mathcal{H}^{(\mathrm{eff})}_{3,D=0}(k)-E_F]_{1,1}&=&[\mathcal{H}^{(\mathrm{eff})}_{3,D=0}(k)-E_F]_{2,2}=\frac{S(4S-3)J^2_{xy}}{4J_z} \frac{(2S-1)J_{xy}\cos k-2J_z}{J_z},\nonumber\\
~[\mathcal{H}^{(\mathrm{eff})}_{3,D=0}(k)-E_F]_{1,1}&=&-\frac{SJ^2_{xy}}{2J_z}\frac{2J_z(4S-1)+J_{xy}S(10S-3) \cos k}{J_z} ,\nonumber\\
~[\mathcal{H}^{(\mathrm{eff})}_{3,D=0}(k)-E_F]_{1,2}&=&z^{-(N+1)}J_{xy}\left[-(2S-1)-\frac{J_{xy}3S(S-1)z^{3}}{2J_z}+\frac{J^2_{xy}S(2S-1)(5S-3)}{4J^2_z}\right],\nonumber\\
~[\mathcal{H}^{(\mathrm{eff})}_{3,D=0}(k)-E_F]_{1,3}&=&\sqrt{S(2S-1)}J_{xy}\left[ -z-\frac{J_{xy}Sz^{-2}}{2J_z}+\frac{J^2_{xy}zS(17S-9)}{8J^2_z}\right],\nonumber\\
~[\mathcal{H}^{(\mathrm{eff})}_{3,D=0}(k)-E_F]_{2,3}&=&z^N\sqrt{S(2S-1)}J_{xy}\left[ -z^{-1}-\frac{J_{xy}Sz^{2}}{2J_z}+\frac{J^2_{xy}S(17S-9)}{8J^2_zz}\right],
\end{eqnarray}
where $z=e^{-ik/3}$.

\end{widetext}

\end{document}